# The effect of Microstructure in Exchange Decoupling of SmCo$_5$/Co bi-layers at low temperatures

Rukshan M. Thantirige[1], Nihar R. Pradhan[2], Mark T. Tuominen[1]

[1] Department of Physics, University of Massachusetts, 666 N. Pleasant St, Amherst, MA, 01002, USA.

[2] National High Magnetic Field Laboratory, 1800 E. Paul Dirac Drive, Tallahassee, FL-32310, USA.

Here, we investigated the influence of grain formation on the magnetization reversal of SmCo$_5$/Co at low temperature. A set of SmCo$_5$/Co bi-layer samples were fabricated under identical conditions on MgO(100) and glass substrates with a Cr underlayer. Analysis of each magnetic layer by an Atomic Force Microscope (AFM) reveals that MgO(100) results small and uniform grain formation of 23 nm in contrast to 57 nm on glass, and x-ray diffraction studies show that the sample on MgO(100) has high crystallinity with SmCo$_5$(11$\bar{2}$0) phase. At room temperature, both samples exhibit good hard magnetic properties with coercivity ($H_C$) of 13.2 kOe and 12.5 kOe, and energy products $(BH)_{max}$ of 14.5 MGOe and 5.3 MGOe for samples on MgO(100) and glass, respectively. Low temperature hysteresis measurements show an exchange decoupling at low temperatures for the sample on glass, and this is due to the formation of large grains on glass that reduces the effective inter-grain exchange coupling between phases.

## I. INTRODUCTION

Exchange-coupled magnets that are composites of soft and hard magnetic materials have been explored since its inception by Kneller and Hawig in 1991 [1]. This new magnet family has many intriguing properties such as high remanence ($M$r), high $(BH)_{max}$, high curie temperatures ($T_C$) and lower cost [1-4], which made them ideal candidates to replace existing hard magnetic materials for a wide range of applications from data storage to energy efficient appliances. In these magnets, $H_C$ and $M$r are determined by the hard and the soft magnetic phases respectively, while effective exchange interaction between phases holds the key to achieving excellent magnetic properties. This exchange interaction across the soft-hard magnet interface depends not only on materials and their physical dimensions but also on the grain size and distribution of each phase, which makes it one of the poorly understood phenomenon in exchange-coupled systems. In thin films, substrate plays a key role in controlling microstructure that can come in the form of epitaxial guidance, epitaxial mismatch or even de-wetting at given processing conditions, making the microstructure is unique to a particular substrate [3, 5-9]. As a result, samples grown on different substrates can have very different magnetization reversal paths, producing different magnetic properties even for the same material combination. As an example, Chowdhury *et. al* .[5] reported that $SmCo_5$/Co bi-layers grown on MgO(110) and Si(100) under identical processing conditions with same layer thickness resulted energy products of 20.1 MGOe and 12.4 MGOe, respectively.

Here, we studied the reversal of exchange spring thin films grown on two different substrates. $SmCo_5$/Co was selected as the exchange-coupled bi-layer and it was grown on MgO(100) and glass substrates under identical conditions. We chose MgO(100) and glass as substrates, as the former promotes uniform grain formation due to epitaxial matching [10-12] while the latter results random grain formation. This allows us to investigate how grain formation determines magnetic properties of this bi-layer sample at various temperatures. Room temperature magnetic measurements confirm spring-exchange behavior with high $H_C$ for both samples, however, low temperature hysteresis measurements show an exchange decoupling like phenomenon for the sample grown on glass. This transformation from single step to two step reversal, below a critical temperature, suggests a weakening of exchange interaction.

## II. SAMPLE PREPARATION

SmCo$_5$/Co exchange spring bi-layer films were fabricated in a DC and RF magnetron sputtering system (Orient 8, AJA Inc.) at high vacuum of $10^{-8}$ Torr on MgO(100) (sample A) and glass (sample B) substrates, which were attached to a rotating stage equipped with a heater. Thickness of each layer was fixed at 60 nm, 30 nm and 7.5 nm for the Cr seed layer, SmCo$_5$ and Co layers, respectively. The nominal thickness of SmCo$_5$ and Co was determined based on highest $H_C$ at room temperature with single-step hysteresis. An alloy target with the proper composition was used to deposit SmCo$_5$, and both the seed and SmCo$_5$ layers were sputter deposited at 500 °C that is adequate enough to induce in-plane hard magnetic properties of SmCo$_5$ layer [13, 14]. After the growth of Cr and SmCo$_5$ layers, samples were allowed to cool for 6 hours before depositing Co layer. This is an important step to minimize inter-diffusion at the SmCo$_5$/Co interface, which can change the composition of the hard phase. A 30 nm Cr layer was deposited onto Co layer to protect magnetic layers from oxidation. All layers were sputter deposited at 4 mTorr in Ar atmosphere and low sputtering powers were used to keep low deposition rates that promotes continues film growth.

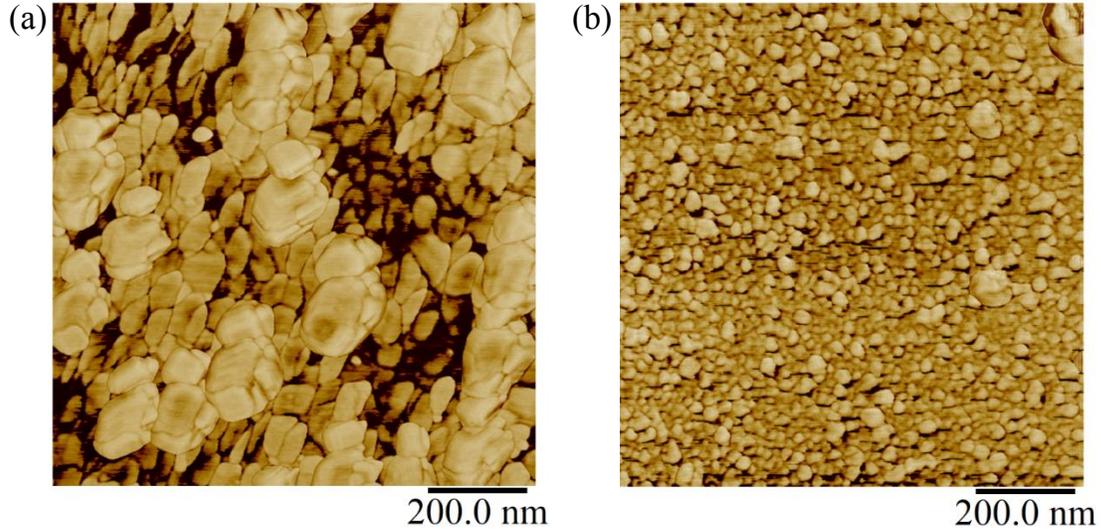

FIG. 1. AFM micrograph of SmCo$_5$ grown at 500 $^0$C on (a) glass and (b) MgO(100) substrates with a 60 nm Cr underlayer. Glass substrate results larger grains with wide size distribution in contrast to grains on MgO(100). The average grain size on glass and MgO(100) are 57 nm and 23 nm, respectively.

## III. RESULTS AND DISCUSSION

The substrates, grain size and distribution of magnetic layers were analyzed by an AFM (Vecco Nanoscope IV, Bruker Inc.) and their crystallinity was investigated by x-ray diffraction with Cu K$_\alpha$ radiation (X'Pert MRD, Panalytical Inc.). Fig. 1 shows the AFM micrographs of a 30 nm SmCo$_5$ film on MgO(100) and glass with a 60 nm Cr layer at 500 ºC. Here, the average grain size of SmCo$_5$ on MgO(100) and glass are 23 nm and 57 nm, respectively, elucidating the formation of smaller SmCo$_5$ grains on MgO(100). Further, as Fig. 1(a) shows, the glass substrate produces SmCo$_5$ grains with wide size distribution in comparison to MgO(100). The peak-to-peak roughness of the SmCo$_5$ film on MgO(100) is 2.8 nm as opposed to 7.0 nm on glass, despite that substrate roughness of MgO(100) and glass are 2.5 nm and 2 nm, respectively. This low roughness of SmCo$_5$ layer on MgO(100) promotes a smoother interface between SmCo$_5$ and Co, and a continuous Co film growth on SmCo$_5$. In contrast, the rough interface between SmCo$_5$ and Co layers may have resulted Co islands as the layer thickness (7.5 nm) is within the range of interface roughness (7.0 nm). In general, graded interfaces produce high $H_C$ as they produce local energy minima and hinder domain wall motion by pinning [15], however, smoother interfaces and smaller grains with large area grain boundaries are considered crucial for effective interfacial exchange coupling [16, 17].

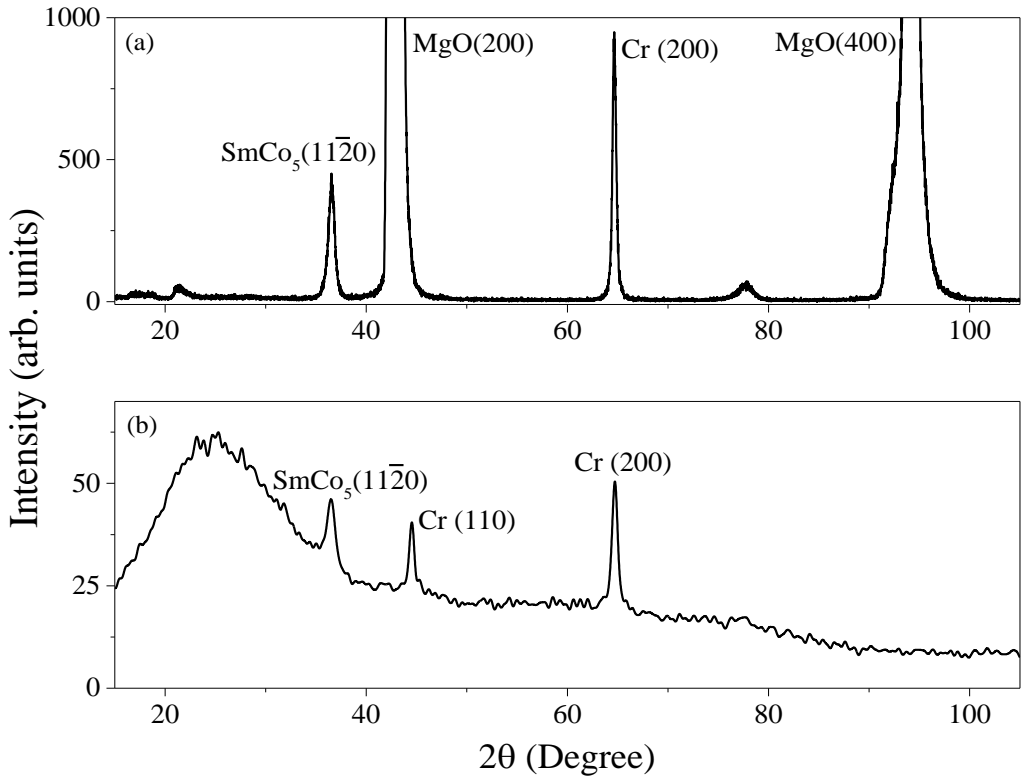

FIG. 2. XRD patterns of Co/SmCo$_5$/Cr trilayers sputtered onto (a) MgO(100), and (b) glass. The growth temperature was set to 500 $^0$C for SmCo$_5$ and Cr layers.

Fig. 2 shows x-ray diffraction patterns of sample A (a) and B (b), respectively. For sample A, strong diffraction peaks corresponds to MgO(200), Cr(200) and SmCo$_5$(11$\bar{2}$0) planes can be observed. This shows that MgO(100) promotes epitaxial growth of Cr(200) that guides the growth of highly textured SmCo$_5$(11$\bar{2}$0) phase. However, for sample B, a peak corresponds to Cr(110) can be observed in addition to Cr(200). The peak intensity of Cr(110) is roughly half of that of the Cr(200) peak. For an isotropic Cr sample at room temperature, the intensity ratio between Cr(200) and Cr(110) phases is about 0.2 [7]. This affirms that the strong signal of Cr(200) phase on glass is driven by high temperature annealing. The formation of SmCo$_5$(11$\bar{2}$0) crystals in sample B, guided by Cr(200) phase, shows that in-plane hard magnetic properties are not unrealistic on glass with a matching buffer layer and proper growth conditions. However, its weak signal assures that most of the SmCo$_5$ are polycrystalline or amorphous. Zhang *et. al.* reported similar observations for SmCo$_5$ samples fabricated on MgO(100) and glass substrates [7].

In-plane magnetic properties were measured by a superconducting quantum interference device (SQUID) magnetometer (MPMS-7T, Quantum Design Inc.) with a maximum field of ±50 kOe. Fig. 3 shows normalized hysteresis of both sample A and B measured at 300 K. The $H_C$ of sample A and B are 13.2 kOe and 12.5 kOe, respectively, suggesting that SmCo$_5$(11$\bar{2}$0) that is aligned in-plane may be responsible for high $H_C$ in both samples. The maximum energy products, $(BH)_{max}$, are 14.5 MGOe and 5.3 MGOe for sample A and B, respectively. This low $(BH)_{max}$ of sample B can be due to low remanence caused by the random orientation of SmCo$_5$ grains as depicted in x-ray diffraction; the weak signal for SmCo$_5$(11$\bar{2}$0). This shows that high $H_C$ does not necessarily guarantee a high $(BH)_{max}$ as random orientation of crystals significantly lowers the effective magnetization and hence the maximum energy product, $(BH)_{max} \propto M^2$.

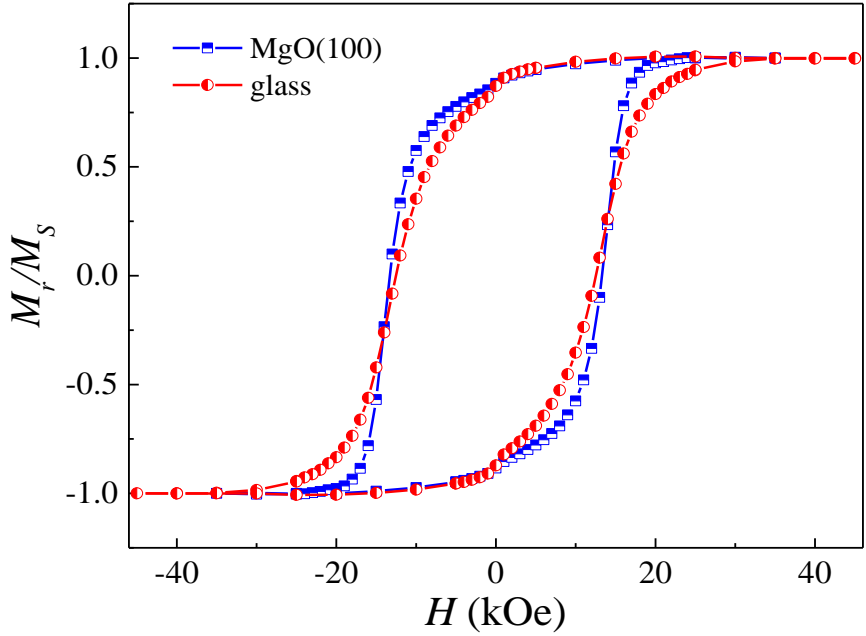

FIG. 3. Normalized room temperature hysteresis curves of $SmCo_5$/Co bilayers on MgO(100) and glass substrates grown at 500º C. The sample on MgO(100) shows higher degree of squareness.

The temperature dependent magnetic properties of these two samples were measured from 300 - 50 K and corresponding hysteresis curves for sample A and B are shown in Fig. 4(a) and 4(b), respectively. For the sample A, $H_C$ rises from 13.2 kOe at 300 K to 23.3 kOe at 50 K but $M_S$ largely remains at $2.42 \times 10^{-4}$ emu except a slight drop at 100 K and 50 K (see Table 1), while preserving the single step reversal behavior. This implies that the exchange coupling between soft and hard phases is preserved at low temperatures, and the increase in $H_C$ can be ascribed to increase in effective magnetocrystalline anisotropy and higher degree of pinning at lower temperatures. However, it should be noted that a slight decrease in the squareness of the hysteresis loop can be observed with lowering the temperature. Zhang *et al*.[18] reported a rise in $H_C$ with decreasing the measuring temperature while preserving the single step reversal for Sm-(Co, Cu)/Fe thin films grown on $SiO_2$, however, that increase took an exponential form, in contrast to an almost linear rise observed in our study.

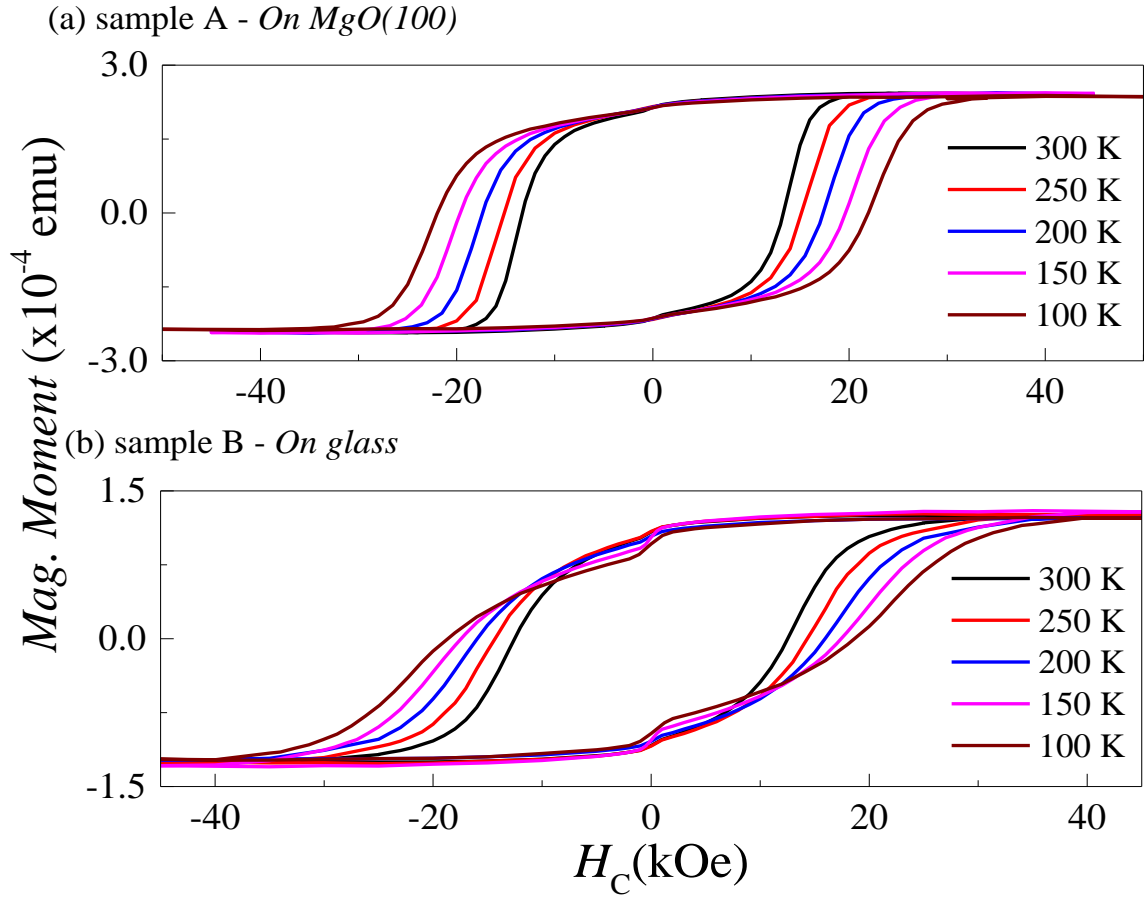

FIG. 4. Low temperature hysteresis curves measured at 300 K (black), 250 K (red), 200 K (blue), 150 K (magenta), 100 K (brown) for the SmCo$_5$/Co sample on (a) MgO(100) and (b) glass. Formation of a 'shoulder' can be seen in hysteresis curves in graph (b) for 150 K and 100 K. Note: 50 K hysteresis is not shown for clarity.

For the sample B, the $H_C$ increases from 12.5 kOe at 300 K to 21.8 kOe at 50 K while the saturation moment shows a random variation with the temperature as shown in Table 1. Although the single step reversal is preserved for higher temperatures (above 150 K) formation of a 'shoulder' in 150 K, 100 K and 50 K (50 K measurement is not shown in Fig. 4) hysteresis curves can be clearly noticed. This transformation from single-step reversal to two-step reversal indicates an exchange decoupling takes place below a critical temperature. This phenomenon has been previously observed and accounted to decoupling of soft and hard phases when lowering the temperature [19, 20] as follows. Based on first principle calculations, effective exchange coupling between soft and hard phases and the single-step reversal require the soft phase to be confined to the size of domain wall width of the hard phase ($\delta_K$) [1-3]. However, $\delta_K$ is governed by the effective anisotropy $K$, as $\delta_K \propto 1/\sqrt{K}$, which increases with decreasing the temperature. This makes $\delta_K$ drops when decreasing the temperature, mandating a smaller soft region to keep the exchange coupling intact at lower temperatures. Since the physical size of the soft region remains unchanged, decreasing the temperature makes these two phases partially or fully decoupled, resulting a two-step reversal process. However, here, we only see such decoupling for sample B despite both samples have identical soft and hard layer thickness and fabricated under identical conditions. Topographic analysis of these samples (Fig. 1) shows that sample A has small and uniform grains in contrast to sample B, which has large grains and inhomogeneous size distribution. This suggests that exchange decoupling observed for sample B is caused by its microstructure. The idea of grain-controlled magnetic properties of thin films can be supported by a number of recent studies that propose smaller grains and uniform distribution favor large inter-grain exchange interactions that enhance the remanence [21-23], and this is in-line with higher remanence and $(BH)_{max}$ of sample A in contrast to sample B. Another key observation in hysteresis of sample B is that there is a cross-over between low temperature (<200 K) and high temperature curves (Fig. 4 (b)). This indicates a change of the reversal mechanism due to the exchange decoupling between hard-soft phases that results a weakening of hard magnetic properties.

TABLE 1. Coercivity ($H_C$), saturation moment and reduced remanence ($M_r/M_S$) for sample A and sample B for temperatures between 300 K – 50 K (extracted from Fig. 4)

| Temperature (K) | Sample A | | | Sample B | | |
|---|---|---|---|---|---|---|
| | $H_C$ (kOe) | Sat. Moment (x$10^{-4}$ emu) | $M_r/M_S$ | $H_C$ (kOe) | Sat. Moment (x$10^{-4}$ emu) | $M_r/M_S$ |
| 300 | 13.2 | 2.42 | 0.88 | 12.5 | 1.24 | 0.87 |
| 250 | 15.1 | 2.41 | 0.90 | 14.3 | 1.26 | 0.87 |
| 200 | 17.3 | 2.42 | 0.89 | 15.9 | 1.23 | 0.88 |
| 150 | 19.6 | 2.42 | 0.88 | 17.4 | 1.28 | 0.80 |
| 100 | 22.0 | 2.38 | 0.90 | 18.6 | 1.23 | 0.80 |
| 50 | 23.3 | 2.39 | 0.89 | 21.8 | 1.29 | 0.79 |

Table 1 depicts the data extracted from Fig. 4(a) and 4(b) at various temperatures from 300 K-50 K and Fig. 5 shows the variation of the reduced remanence ($M_r/M_S$) and coercivity ($H_C$) with the temperature. As shown in Fig. 5(b), $H_C$ increases almost linearly with the temperature with slightly different gradients of -0.042 and -0.035 with extrapolated coercivity of 25.76 kOe and 22.84 kOe at 0 K for sample A and B, respectively. This steep increase in $H_C$ when lowering the temperature for sample A can be associated with large concentration of domain wall boundaries resulted by smaller grains, compared to those of sample B. Further, Fig 5(a) shows that $M_r/M_S$ has no significant temperature dependence for sample A, however, a sharp drop from 0.88 to 0.80 for sample B can be observed when the measuring temperature is reduced from 200 K to 150 K. This is a result of weakened exchange coupling between soft and hard magnetic grains when the measuring temperature is reduced, which essentially reduces the remanence.

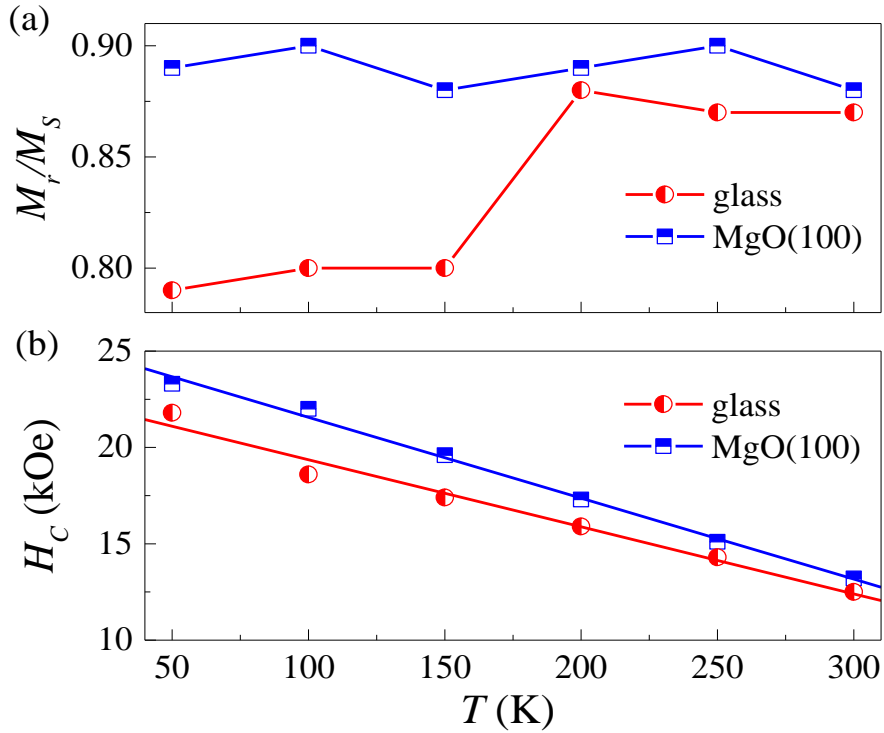

FIG. 5. Variation of (a) reduced remanence ($M_r/M_S$), and (b) coercivity ($H_C$) with temperature for sample A (blue) and sample B (red). $H_C$ was fitted to linear functions with corresponding gradient of -0.042 and -0.035 for sample A and sample B, respectively.

## IV. CONCLUSIONS

In this work, we investigated magnetization reversal of exchange-coupled magnetic thin films fabricated on single crystal MgO(100) and amorphous glass substrates with a 60 nm Cr seed layer. X-ray diffraction studies show that in-plane hard magnetic properties are due to the formation of twisted-crystalline SmCo$_5$ (11$\bar{2}$0) phase, guided by the Cr (200) seed layer. AFM measurements of each magnetic layer reveals that MgO(100) induced small and uniform grains of 23 nm in contrast to larger grains (57 nm) on glass with a random size distribution. Room temperature hysteresis measurements show that both samples exhibit good hard magnetic properties with high coercivity but $(BH)_{max}$ of the sample on glass (5.3 MGOe) is almost 1/3 of that of the sample on MgO(100) due to random orientation of crystals that lowers the effective magnetization. Hysteresis measurements at lower temperatures reveal an exchange decoupling like phenomenon only for the sample on glass. We believe that this decoupling is induced by the microstructure, as large magnetic grains on glass substrate reduce the effective inter-grain exchange coupling between soft and hard magnetic phases, which is critical against rising anisotropy when reducing the temperature.

ACKNOWLEDGMENTS

This work was supported by NSF grants DMR-1208042 and NSEC Center for Hierarchical Manufacturing-CMMI-1025020.